\documentclass[a4paper,fleqn,usenatbib]{mnras}
\usepackage{graphicx}
\usepackage{amssymb}
\usepackage{amsmath}
\usepackage{times} 
\usepackage{multirow}
\usepackage{url}

\begin{document}

\title[Baryon acoustic oscillations signature in the three-point angular correlation]
{Baryon acoustic oscillations signature in the three-point angular correlation function from 
the SDSS-DR12 quasar survey}

\author[E. de Carvalho, A.  Bernui, H. S. Xavier, C. P. Novaes]
{E. de Carvalho$^{1,2}$\thanks{e-mail: edilsonfilho@on.br}
A. Bernui,$^{1}$ 
H. S. Xavier,$^{3}$ 
C. P. Novaes$^{1,4}$ \\ 
$^{1}$Observat\'orio Nacional, Rua General Jos\'e Cristino 77, 
          S\~ao Crist\'ov\~ao, 20921-400 Rio de Janeiro, RJ, Brazil \\
$^{2}$Centro de Estudos Superiores de Tabatinga, Universidade do Estado do Amazonas, 
69640-000, Tabatinga, AM, Brazil \\
$^{3}$Instituto de Astronomia, Geof\'isica e Ci\^encias Atmosf\'ericas, Universidade de S\~ao 
Paulo, Rua do Mat\~ao, 1226, 05508-090, S\~ao Paulo - SP, Brazil\\
$^{4}$ Instituto de F\'isica, Universidade de S\~ao Paulo, Rua do Mat\~ao trav. R 187, 05508-090, S\~ao Paulo - SP, Brazil
}

\date{\today}
\maketitle

\begin{abstract}

\noindent
The clustering properties of the Universe at large-scales are currently being probed at various 
redshifts through several cosmological tracers and with diverse statistical estimators. 
Here we use the three-point angular correlation function (3PACF) to probe the baryon acoustic 
oscillation (BAO) features in the quasars catalogue from the twelfth data release of the Sloan 
Digital Sky Survey, with mean redshift $\overline{z} = 2.225$, detecting the BAO imprint with a 
statistical significance of $2.9 \,\sigma$, obtained using lognormal mocks. 
Following a quasi model-independent approach for the 3PACF, we find the BAO transversal 
signature for triangles with sides $\theta_1=1.0^\circ$ and $\theta_2=1.5^\circ$ and the angle 
between them of 
$\alpha=1.59 \pm 0.17$ rad, 
a value that corresponds to the angular BAO scale 
$\theta_{\mbox{\sc bao}}=1.82^{\circ} \pm 0.21^{\circ}$, in 
excellent agreement with the value found in a recent work 
($\theta_{\mbox{\sc bao}}=1.77^{\circ} \pm 0.31^{\circ}$) applying the 2PACF to similar data. 
Moreover, we performed two type of tests: one to confirm the robustness of the BAO signal 
in the 3PACF through random displacements in the dataset, 
and the other to verify the suitability of our random samples, a null test that in fact does not 
show any signature that could bias our results. 
\end{abstract}

\maketitle

\section{\label{sec1} Introduction}

Studies of the large-scale structure (LSS) have revealed properties of the Universe which confirm 
the $\Lambda$CDM hierarchical scenario for galaxy formation and cosmic 
evolution~\citep{Peacock,Springel,Piattella}. 
The information about LSS has been accessed mainly using the $n$-point correlation function 
statistics~\citep{Peebles01,PeeblesGroth75,GrothPeebles77}. 
Thereby, the two-point correlation function (2PCF) was extensively employed to search for the 
Baryon Acoustic Oscillations (BAO) imprint in the galaxy and quasar 
surveys~\citep{PeeblesYu,SZ,BE1984,Cole05,Eisenstein05,Paris,marra2018}. 
The next order statistics, the three-point correlation function (3PCF), has been used to probe 
the non-Gaussian features expected in the galaxy 
distribution~\citep{Frieman99,Slepian17a,Slepian17b}, and to confirm the predictions of non-linear 
cosmological perturbation theory \citep[see, e.g.,][for a review]{Bernardeau02}.

The 3PCF is being also used to confirm the BAO features; the first analyses of this type 
were done by~\citet{Gaztanaga09} with the sixth and seventh data releases (DR) from the Sloan 
Digital Sky Survey (SDSS), where they found the BAO signature at $\sim 100 \mbox{Mpc}/h$. 
Recently,~\citet{Slepian17a,Slepian17b} detected the BAO signal in the 3PCF with 
$4.5\sigma$ statistical significance using the SDSS DR12 galaxy sample. 
Many of the reported works assume a fiducial cosmology to calculate the three-dimensional (3D) 
comoving distances between the pairs of cosmic objects that form a triangle configuration to finally 
compute the 3PCF. 
Some of these works, e.g.~\cite{Frieman99,Jing04,McBride10}, perform their analyses in the 
projected space~\citep{Davis-Peebles}. 
To minimize the impact of redshift distortions, they first calculate the 2PCF as a function of 
two coordinates: the redshift space distance into line-of-sight, $\pi$, and projected separation, 
$r_p$, such that $(\pi^2 + r_p^2)^{1/2}$ is the observational distance in redshift space. 
Because the anisotropic redshift space distortion is primarily contained in the $\pi$ coordinate, 
they integrate along this coordinate resulting in the projected 2PCF, and then the projected 3PCF 
is obtained through analogous definitions \citep[see, e.g.,][for a review]{McBride10}.

In 2011,~\citet{Sanchez11} proposed an approach to calculate the two-point angular correlation 
function (2PACF) in a quasi model-independent way. 
This methodology was then applied by~\citet{Carnero11} to study the angular BAO signature of 
the DR7 SDSS sample of luminous red galaxies. 
After that, the 2PACF has been applied to several datasets to investigate the BAO signal at 
different redshifts~\citep{Gabriela1,Gabriela2,Salazar17,Abbott17,Crocce17,Edilson18}. 
Here, we extend to the three-point statistics the approach proposed by~\citet{Sanchez11} and 
perform for the first time analyses of the three-point angular correlation function (3PACF) based 
only in the sky angular separation of SDSS quasars located in a thin redshift shell, with mean redshift 
$\overline{z} = 2.225$. 
We successfully confirm the BAO transversal signature at the same angular position already 
found in a recent work analyzing these data with the 2PACF~\citep{Edilson18}; from now on 
this reference is termed EdC18.


The main motivations to perform two-dimensional (2D) BAO analyses, instead of 
the 3D approach, are the following. 
Differently from the 3D case where one needs to assume a fiducial cosmology to calculate the 
comoving distances between pairs of objects in order to construct the 2PCF, 
in 2D analyses one only uses the angular coordinates, given by the survey catalogue, 
to calculate angular distances between pairs to search for the BAO features in the 2PACF and 
3PACF. 
An advantage of such model-independent approach is that their results can be combined with 
other model-independent (or weakly model-dependent) data to impose restrictions on cosmological 
models or parameters, or simply to compare results obtained in a 3D approach. 
One can also perform 2D analyses in several non-correlated thin redshift bins to obtain the best-fit 
angular diameter distance $D_A(z;r_s)$, to be used in cosmological model or parameter analyses 
as done by \cite{Edilson18,Gabriela1,Carnero11,Sanchez11}.
In addition, if the main target in BAO analyses is a statistically significant measurement of the BAO 
signature, another advantage is that some undesired phenomena that affects such measure in 
3D are minimal or negligible in 2D analyses considering data in thin redshift bins (e.g. the 
redshift space distortions).

We organise this work as follows. 
Section~\ref{sec2} gives the details of the quasars, the random, and the mock catalogues 
employed in the analyses; and the angular correlation function estimators applied to these datasets are 
presented in section~\ref{sec3}. 
The data analyses and results are discussed in section~\ref{sec4}, while in section~\ref{sec5} we 
summarise our conclusions.

\section{The data, random, and mock catalogues}\label{sec2} 

\subsection{The quasars and randoms dataset}
The data used is part of the twelfth public Data Release Quasar catalogue (DR12Q), from the 
SDSS-III~\citep{Eisenstein2011}%
\footnote{\url{www.sdss.org/dr12/algorithms/boss-dr12-quasar-catalog/}}. 
The DR12Q sample contains $297,301$ quasars from the Baryon Oscillation Spectroscopic Survey
~\citep[BOSS;][]{Dawson2013}, among which $184,101$ have $z \ge 2.15$, covering a total sky 
area of $9,376 \, \mbox{deg}^{2}$. 
The full sample has been spectroscopically confirmed based on a visual inspection of the spectra 
of each quasar. 
The SDSS-III/BOSS limiting magnitudes for quasar target selection are $r \le 21.85$ or 
$g \le 22$~\citep{Paris}. 
The main challenge faced in the quasar BOSS survey was to obtain a high number density sample, 
satisfying the proposed minimum threshold of 15 quasars per square degree~\citep{Paris}. 
This sample is dense enough to perform 2D analyses in thin redshift bins. 

In EdC18 we performed a detailed evaluation of the signal-to-noise ratio to select 
the quasars data for BAO analyses. 
As a result, we selected a sample of quasars in the thin shell $z \in [2.20, 2.25]$, with width 
$\delta z = 0.05$, containing a total of $13,980$ quasars distributed between the north and south 
Galactic hemispheres (in EdC18 we consider only the data in the north Galactic region). 
The number density of this dataset is large enough to measure the angular BAO signature with 
a good statistical significance using the 3PACF. 

The random catalogues are used to extract the BAO features from the data, for this they must 
share common properties as those observed in the quasar catalogue. 
The random samples were generated according to the procedure described in EdC18; 
for the present analyses we produced 150 random sets with equal number of objects, 
homogeneously distributed in the same sky region as the quasars catalogue; 
50 of these sets were used for the 2-point and 3-point correlation functions statistics, while 
we employed the other 100 sets for the null test analyses.

\subsection{The Mocks}\label{mocks}

The mock quasar catalogues used in this work are full-sky lognormal realizations created with 
the~\texttt{FLASK} code\footnote{\url{http://www.astro.iag.usp.br/\~{ }flask}} \citep{Xavier16}. 
To generate such mocks in a single redshift shell, we provided as input: the expected projected 
number density of quasars of 1.49 $\mathrm{deg^{-2}}$ (the same as in BOSS data); and a 
fiducial angular power spectrum $C_\ell$ computed 
with~\texttt{CAMBsources}\footnote{\url{http://camb.info/sources}} \citep{Challinor11} 
for a top-hat redshift bin ($2.20 < z < 2.25$), assuming a quasar linear bias of 4.25, the $\mathrm{\Lambda CDM}$ cosmological parameters measured by Planck \citep{Ade15} and 
a minimal Neutrino contribution (effective number of neutrinos $N_{\mathrm{eff}}=3.046$ and 
sum of masses $\Sigma m_\nu=0.06\mathrm{eV}$). 
All $C_\ell$ contributions available in \texttt{CAMBsources} (e.g. lensing, redshift space 
distortions and non-linear clustering) were included. The shift parameter $\lambda$ of the 
lognormal probability distribution ($-\lambda$ is the minimum value attained by the quasar 
density contrast) was set to 1.

Once the mean number density, the shift parameter and the angular power spectra are defined 
in the lognormal model, all other statistical properties are set in accordance, including the 
3PACF~\citep{Xavier16}. 
We adopted an angular resolution for the mocks of $\sim 0.06 \,\mathrm{deg}$, set by the 
\texttt{Healpix}\footnote{\url{http://healpix.sourceforge.net}} \citep{Gorski05} parameter 
$N_{\mathrm{side}}=1,024$. 
On scales smaller than this, the mock quasars are distributed homogeneously (their distribution 
inside a pixel is random). 
In conformity with the simulation's resolution, we band-limited the realizations to 
$\ell_{\mathrm{max}} = 2,560$. 
A total of 200 full-sky mock catalogues were produced for our analyses.

\section{The angular correlation functions}\label{sec3} 

Many BAO analyses assume a fiducial cosmology to compute the comoving distance among 
pairs, then the characteristic scale is found through the 2PCF, 
and similarly for the computation of the 3PCF. 
We are interested in the transversal BAO signal, for this we use 
the angular version of this estimator, i.e., the two-point angular 
correlation function (2PACF) and the next order, the 3PACF, that will be applied to the quasars 
data in a thin redshift shell.

\subsection{The three-point angular correlation function}

The 3PCF is a complementary tool to characterize the clustering of cosmological tracers like 
galaxies, quasars, etc. For other applications of the 3PCF see, 
e.g.,~\cite{Fry82,Jing04,Gaztanaga05II,Kulkarni07,%
McBride10,Marin11,Marin13}; for alternative statistical tools and clustering analyses see, 
e.g.,~\cite{Camila1,Camila2,Camila3,GAM,GAM2}. 
Basically, the 3PCF compares the number of triplets of cosmic objects from a dataset that 
form a triangle configuration, to be called $DDD$, with respect to the number of triplets from 
a simulated random set of data, termed $RRR$. 

Let us start briefly reviewing the basics of the 2PCF. 
This statistical tool measures the excess probability over a random dataset of finding pairs of 
cosmic objects from a given catalog. 
It has been used for many applications in astrophysical problems~\citep{Peebles01,%
BFW,Salazar14,Avila,Avila2}. 
To calculate the 2PCF, the widely used estimator is the Landy-Szalay (LS) 
estimator~\citep{Landy-Szalay}, which has a better performance when compared with other 
estimators~\citep{Kerscher} because it results in the smallest deviations for a given 
cumulative probability, besides having minimal variance and no bias. 
The LS estimator is defined by 
\begin{eqnarray}\label{2PCF}
\xi(s) \equiv \frac{DD(s) - 2DR(s) + RR(s)}{RR(s)} \, ,
\end{eqnarray} 
where $DD(s)$, $RR(s)$, $DR(s)$ are the normalized pair counts between data-data, 
random-random and data-random objects, respectively, where the pairs are separated by the 
comoving distance 
$s$~\citep{Landy-Szalay,Sanchez11}. 
In addition, to estimate the 3PCF we consider the Szapudi-Szalay, SS, estimator~\citep{SS}, 
which is a general extension for all $n$-point correlation functions in 3D. 
For the case of the 3PCF, $n=3$, the SS estimator assumes the form 
        \begin{eqnarray}\label{3PCF}
	\zeta(S) \equiv \frac{DDD(S) - 3DDR(S) + 3DRR(S) - RRR(S)}{RRR(S)} \, ,
	\end{eqnarray}
where the $DDD$ and the other terms are all normalized triplet counts such that three 
cosmic objects form a triangle of sides given by the triplet 
$S=\lbrace s_{12},s_{23},s_{31}\rbrace$, where $s_{12}$ is the comoving distance between the 
objects 1 and 2 and so on~\citep[see][]{Marin11}.

The correlation functions that explore the clustering of objects in the 3D space need to assume 
a fiducial cosmological model to calculate the 3D distances first and then the comoving 
distance $s$ between pairs of cosmic objects. 
However, using the angular version, that is the 2PACF and 3PACF, one can minimize this 
model dependence, by considering just angular distances in the transversal plane (actually, 
a thin shell) to the line of sight. 
In this case, the data is located in a thin redshift bin, and the 2PACF measures the 
transversal BAO signature. 
The angular version of the LS estimator, $w(\theta)$, for data in a thin redshift bin 
with mean redshift $\overline{z}$, is given by~\citep[see, e.g.,][]{Sanchez11,Carnero11,Gabriela1} 
\begin{eqnarray}\label{2PACF}
w(\theta) \equiv \frac{DD(\theta) - 2DR(\theta) + RR(\theta)}{RR(\theta)} \, ,
\end{eqnarray}
where $\theta$ is the angular separation between any pair in the data and/or in the random 
sample. 
Analogously, the angular version of the SS estimator~\citep{PeeblesGroth75,%
Frieman99,Materne,Cardenas} for the 3PACF 
involves 3 variables that define the triangle formed by 3 cosmic objects, is 
\begin{eqnarray}\label{3PACF}			    
\!\!W(\Theta) \!\equiv \frac{\!DDD(\Theta) \!-\! 3DDR(\Theta) \!+\! 3DRR(\Theta) \!-\! 
RRR(\Theta)}{RRR(\Theta)} ,
\end{eqnarray}
where $\Theta$ represents the triplet of angular distances $\{ \theta_1,\theta_2,\theta_3 \}$ 
of the triangle. 
Specifically, $\theta_1$ ($\theta_{2}$, $\theta_{3}$) is the angular 
distance between the cosmic objects number 2 (3, 1) and number 3 (1, 2).

One can introduce the {\em reduced} 3PACF, defined by~\citet{GrothPeebles77} as 
\begin{eqnarray} \label{q3PACF}
q(\Theta) = \frac{W(\Theta)}{\,w_{1}w_{2} + w_{2}w_{3} + w_{1}w_{3}\,} \, ,
\end{eqnarray}

\vspace{0.15cm}
\noindent
where $w_{i} \equiv w(\theta_{i}), \,i=1, 2, 3$, with $\theta_{i}$ as explained above. 
According to~\cite{Edilson18}, the angular BAO scale is $1.77^\circ\pm 0.31^\circ$ for this 
redshift bin $z \in [2.20,2.25]$, therefore we have an expectation for the triangle 
configuration and its scale. Furthermore, we choose to analyze $q$ it instead of $W$ because, as noted by~\cite{Marin13}, it appears to be more suitable to study the shape dependence of matter clustering. 
Besides this advantage, 
\cite{fry1994} have shown that the non-linear bias affects just the amplitude of 
the reduced 3PACF but not the triangle shape considered in the analysis~\citep[see also][]{Zheng04,Gaztanaga09}. 

Additionally, one can parametrize the triplet configurations in the following way. 
One first fixes the values $\theta_{1}$ (the angular distance between objects 2 and 3) and 
$\theta_{2}$ (the angular distance between objects 3 and 1), and then calculates the function 
$q(\Theta) = q(\alpha[\theta_{3}])$, for $\alpha \in \left[0^\circ, 180^\circ\right]$, which is the 
angle formed by the sides 2-3 and 3-1 of the triangle 1-2-3
~\citep[the Figure 1 from][illustrates the meaning of $\alpha$]{Gaztanaga05II}:
\begin{eqnarray}\label{eq6}
\cos{\alpha} = \frac{\,\theta_{1}^{2} + \theta_{2}^{2} - \theta_{3}^{2}\,}{2\,\theta_{1}\,\theta_{2}} \, . 
\end{eqnarray}
For $\alpha = 0^\circ$ the configuration is termed {\em collapsed} triangle and the size of the 
third side of the triangle is $\theta_{3} = |\theta_{2} - \theta_{1}|$. 
For the case $\alpha = 180^\circ$, termed {\em elongated} triangle, the third side is 
$\theta_{3} = \theta_{2} + \theta_{1}$~\citep{Gaztanaga05I,McBride10}. 

To find the angular scale of the BAO bump in the {\em reduced} 3PACF, $q(\alpha)$, we follow 
the approach of~\citet{Sanchez11} based on an empirical parametrization of $q(\alpha)$ which 
consists of a quadratic function to describe the overall shape, in some works called the `U'-form, 
plus a Gaussian function to describe the BAO bump 
\begin{eqnarray}
q(\alpha) = m + n\,\alpha + p\,\alpha^2 + C\exp^{-(\alpha-\alpha_{\mbox{\tiny\rm FIT}})^2 / 
2 \sigma_{\mbox{\tiny\rm FIT}}^{2}} \, ,
\label{adjust_q3PACF}	
\end{eqnarray}
where $m$, $n$, $p$, $C$, $\alpha_{\,\mbox{\footnotesize\sc fit}}$, 
and $\sigma_{\,\mbox{\footnotesize\sc fit}}$ are free parameters. 
The best-fit of the {\em reduced} 3PACF obtained with this expression provides 
$C$, $\alpha_{\,\mbox{\footnotesize\sc fit}}$ and $\sigma_{\,\mbox{\footnotesize\sc fit}}$ 
which describes the BAO signal shape.
The parameters $m$, $n$, and $p$ control both the amplitude and the form of the parabola.

\section{Data analyses and Results}\label{sec4}

In this section, we perform the analyses that lead us to a robust measurement of the angular BAO 
scale in the DR12 SDSS quasar catalog through the {\em reduced} 3PACF statistic, $q(\alpha)$. 
This work extends the analyses done in EdC18, where the 
2PACF was applied to the north Galactic hemisphere data of the DR12 quasar catalog, in the 
same redshift bin as here, finding a BAO signal at $1.77^{\circ}\pm 0.31^{\circ}$ with statistical 
significance of $2.12 \,\sigma$.

\subsection{The reduced 3PACF results}\label{bin-size}

\begin{figure}
\mbox{\hspace{-0.4cm}
\includegraphics[width=9.6cm, height=7.5cm]{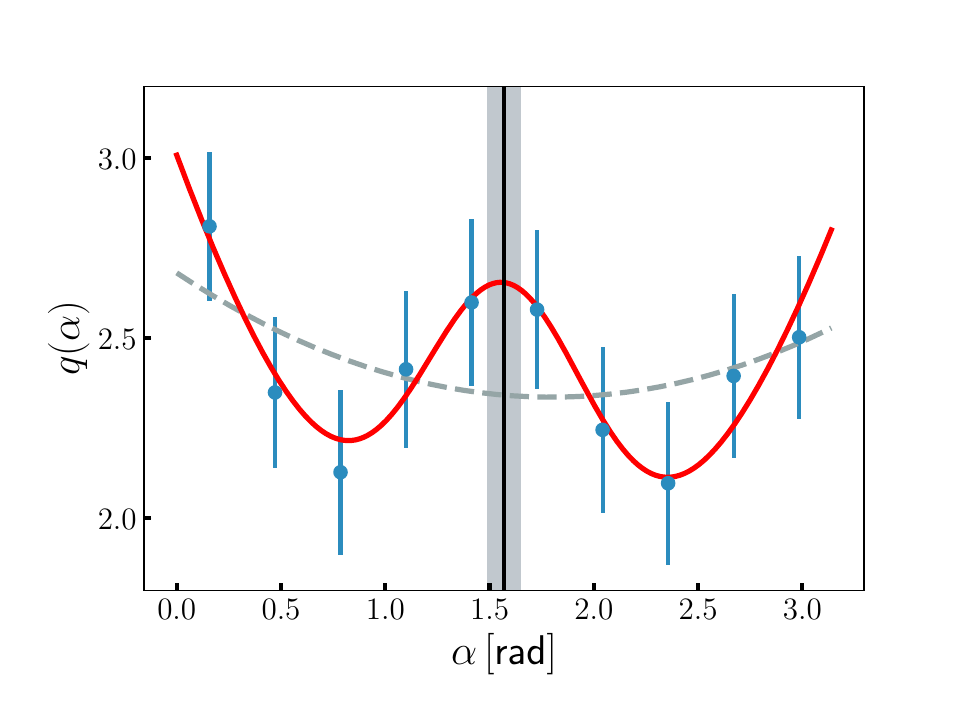}
}
\vspace{-0.4cm}
\caption{The reduced 3PACF (dots), $q(\alpha)$, calculated from the quasars sample 
DR12 -- SDSS with $\overline{z} = 2.225$, for fixed values $\theta_{1} = 1.0^{\circ}$ and 
$\theta_{2} = 1.5^{\circ}$. 
The best-fit of these data is $\alpha_{\mbox{\sc fit}} = 1.57 \pm 0.081_{(stat)}$ rad 
(vertical line; error bar repesented by the gray region), 
where the continuous line was obtained using the equation~(\ref{adjust_q3PACF}) 
considering $N_{b} = 10$ bins. 
The dashed line corresponds to the best-fit curve for the \textit{non}-BAO signal case, 
i.e., $C = 0$ in equation~(\ref{adjust_q3PACF}).} 
\label{fig1}
\end{figure}

Now we shall compute the {\em reduced} 3PACF, $q(\Theta)$, given in equation (\ref{q3PACF}), 
but before we need to calculate the functions $W(\Theta)$ and $w_i$ for $i=1,2,3$, using the 
equations (\ref{2PACF}) and (\ref{3PACF}) for triangle configurations with fixed 
$\theta_{1} = 1.0^{\circ}$ and $\theta_{2} = 1.5^{\circ}$. 
To compute these functions we use 50 random samples, generated according to the procedure 
described in EdC18. 
In this way, we obtain the 3-point angular correlations for our quasars catalogue, in the redshift 
range $z \in [2.20, 2.25]$, with mean redshift $\overline{z}=2.225$, as shown in the Figure~\ref{fig1}. 

The procedure for computing  the reduced 3PACF consists on calculating 
the 2PACF and 3PACF of the quasar catalogue using each random sample. 
The final results are obtained
as the mean over the 50 sets of each $w(\theta)$ and $W(\alpha)$ data points
for every $\theta$ and $\alpha/\pi$ bins. 
Finally, we used the equation (\ref{q3PACF}) to obtain the reduced 3PACF.
The error bars shown in Figure~\ref{fig1} were obtained from the quasar mocks' covariance 
matrix (estimated according to Sec.~\ref{ss-covmatrix}) as the square root 
of the main diagonal for each function, namely $w(\theta)$, $W(\alpha)$, and $q(\alpha)$ 
(see Figure~\ref{fig2}). 

The binning choice, mainly for the 3PACF case, is a compromise that has a strong impact in 
the signal to noise ratio \citep[see][and references therein]{Marin11}. 
Besides that, triplet configurations depend on the angular separation $\theta$ between pairs 
and, to form a reasonable number of triplets, one must choose a value for $\Delta\theta$, 
which will define the resolution of the results, in such a way that we actually do not have 
exact values of $\theta_{1}$ and $\theta_{2}$, but  bins of 
$\theta_{1} \pm \Delta\theta$ and $\theta_{2} \pm \Delta\theta$. 
A low resolution implies a small number of triplets by bin and a small signal-to-noise ratio. 
For our analyses, after several tests, we have chosen $\Delta\theta = 0.15^\circ$ which allow 
us to find a significant number of triplet configurations providing a BAO signature with a good 
statistical significance, as we shall see. 

Note that the reduced 3PACF was estimated for equally spaced bins of $\alpha / \pi$ in the range 
$0.0 \leq \alpha / \pi \leq {1.0}$, in a total of $N_{b} = 10$ bins. 
Then, to extract the BAO features we fit equation~(\ref{adjust_q3PACF}) to the reduced 
3PACF data using the covariance matrix obtained from quasar mocks (see Figure~\ref{fig2}).  
The BAO bump is identified at the position 
$\alpha_{\mbox{\footnotesize\sc fit}} = 1.57 \pm 0.081_{(stat)}$ rad or 
$\alpha_{\mbox{\footnotesize\sc fit}} = 89.89^{\circ} \pm 4.6^{\circ}_{(stat)}$. 

The statistical error, denoted {\it stat}, was obtained in the following way. 
We produce 10,\,000 synthetic $q(\alpha)$ datasets and extract the U-form and BAO bump 
parameters by fitting them according to the empirical parametrization $q(\alpha)$, given by 
equation~(\ref{adjust_q3PACF}). 
Each synthetic dataset was generated by setting the measured $q(\alpha)$ as the true one 
and adding to it Gaussian random errors according to the measured covariance matrix. 
Figure~\ref{fig3} shows the histogram of the recovered BAO bumps, $\alpha_{\mathrm{FIT}}$, 
from these 10,\,000 synthetic realisations, 
whose standard deviation, $\sigma_{stat} = 0.081$ rad, 
gives a measure of the statistical uncertainty in our procedure.
 
The systematic error, denoted {\it sys}, will be calculated in detail in another section. 
According to the values considered for $\theta_{1}, \,\theta_{2}$ above and using the 
equation~(\ref{eq6}) the angle $\alpha = \alpha_{\mbox{\footnotesize\sc fit}}$ 
corresponds to 
$\theta_{\mbox{\footnotesize\sc fit}} \equiv \theta_3 = 1.80^\circ$.

\subsection{The Covariance Matrix estimation}\label{ss-covmatrix}

To estimate the covariance matrix and the significance of our results we have used a sample 
with $N=200$ quasar mocks described above (see the subsection~\ref{mocks}). 
For each mock, we extract the information about the 2PACF and the 3PACF and finally 
calculate the reduced 3PACF, $q(\alpha)$. 
The covariance matrix for $w(\theta)$, $W(\Theta)$ and $q(\Theta)$ was estimated using 
the following expression~\citep[see][]{Gaztanaga09}: 
\begin{eqnarray}\label{covmateq}
	Cov_{ij}= \frac{1}{N}\sum_{k=1}^{N}\left[x_k(i)-\hat{x}(i)\right]\left[x_k(j)-\hat{x}(j)\right] \, .
\end{eqnarray} 
Here, the $x_k(i)$ represents the statistic used [i.e., $w(\theta)$, $W(\Theta)$, or $q(\Theta)$] 
in the bin $i$ for each mock $k$, and the $\hat{x}(i)$ is the mean value for this statistic over 
the 200 mock samples in that bin. 

\begin{figure}
\mbox{\hspace{-0.2cm}
\includegraphics[width=10.cm, height=9.1cm]{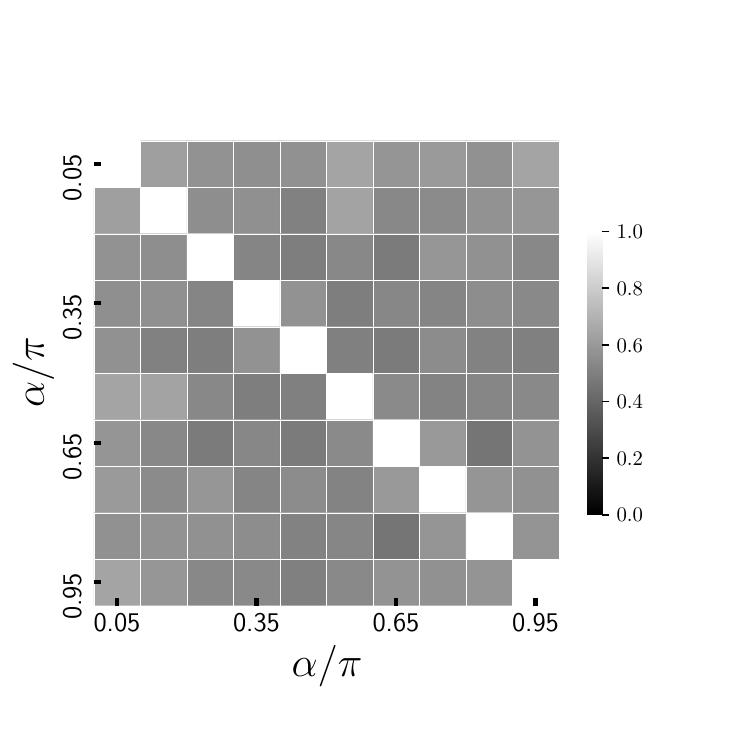}
}
\vspace{-0.6cm}
\caption{The correlation matrix for the reduced 3PACF, $q(\alpha)$, obtained from 200 quasar 
mocks (see section~\ref{sec2} for details of how these mocks were produced, and 
subsection~\ref{ss-covmatrix} for the matrix calculation).} 
\label{fig2}
\end{figure}

\begin{figure}
\mbox{\hspace{-0.3cm}
\includegraphics[width=0.5\textwidth]{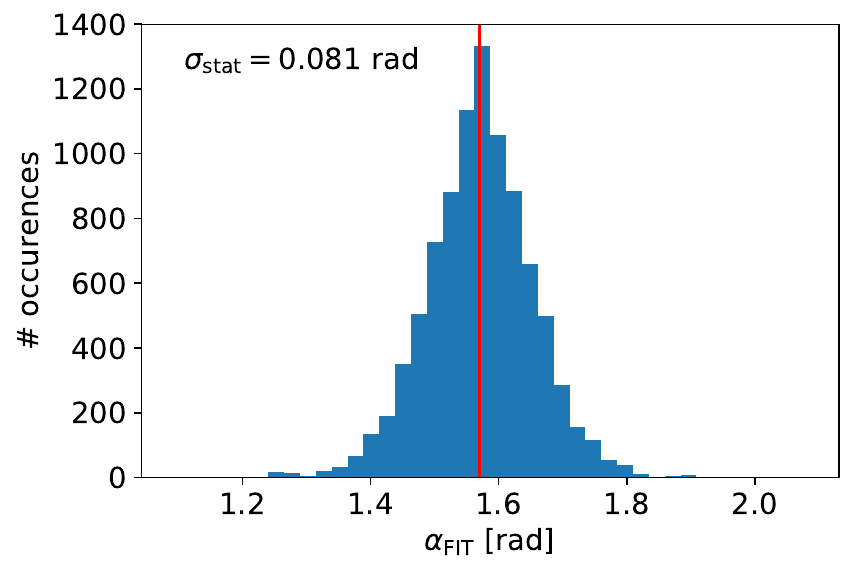}}
\vspace{-.4cm}
\caption{Histogram of best-fitting $\alpha_{\mathrm{FIT}}$ for 10,000 synthetic realisations of 
$q(\alpha)$, assuming the model given by eq. (7), the parameters given by Table 1 
($C \neq 0$) and the Gaussian random errors generated from the covariance 
matrix estimated from the data. 
The red vertical line shows the true value, and the standard deviation of the recovered 
$\alpha_{\mathrm{FIT}}$ is $\sigma_{stat} = 0.081$ rad. 
}
\label{fig3}
\end{figure}

The error of $x(i)$ is the square root of the main diagonal, $\delta x(i)= \sqrt{Cov_{ii}}$. 
We show the covariance matrix in Figure~\ref{fig2} for the case of the reduced 3PACF, 
$q(\alpha)$.

The statistical significance of the BAO signal measurement is obtained through the $\chi^2$ methodology, using the inverse of the covariance matrix as

\begin{eqnarray}\label{covmatrix}
	\chi^2(\alpha) = \left[q(\alpha) - q^{\mbox{\small\sc f\/it}}(\alpha)\right]^{T}Cov^{-1} 
	\left[q(\alpha) - q^{\mbox{\small\sc f\/it}}(\alpha)\right] \, .
\end{eqnarray} 
The symbols $\left[\,\right]$ and $\left[\,\right]^{T}$ represent column vectors and row vectors, respectively. 
We adjusted the parameters of equation (\ref{adjust_q3PACF}), based in the minimum $\chi^2$ method, for two cases:
considering $C$ as  free parameter, $C\ne0$ ($\chi^{2}_{min}=2.24$), and imposing $C=0$ ($\chi^{2}_{min}=16.00$), the latter representing the \textit{non}-BAO case (see Figure~\ref{fig1}).
Table~\ref{table1} shows the best-fit parameters for each case.

As a result, the best-fit of the {\it non}-BAO case (7 degrees of freedom, {\it dof}), compared to the BAO case (4 {\it dof}), is disfavoured by $\Delta\chi^2 = 13.76$
($7-4=3$ \textit{dof}). 
Therefore, our detection of the BAO signal has a significance of $2.9\,\sigma$.

It is worth to notice that this statistical significance depends on the covariance matrix derived from the lognormal simulations, which could underestimate the error bars and, consequently, overestimate the statistical significance. As a comparison, we have used the Jackknife approach to extract the covariance matrix and finding a significance of $1.95\,\sigma$. However, \cite{Norberg} have shown that the Jackknife approach could overestimate the error bars, underestimating the statistical significance. 

\linespread{1.25}
\begin{table}
\centering
\begin{tabular}{c|c|c}
\hline
Parameters \,\,&\,\, Eq.~\ref{adjust_q3PACF} ($C=0$) & Eq.~\ref{adjust_q3PACF} ($C\neq0$) \\
\hline
$m$            & \,\,2.68$\pm$0.31  & \,\,3.01$\pm$0.19 \\
$n$            & \,\,-0.38$\pm$0.30 & \,\,-1.56$\pm$0.54 \\ 
$p$            & \,\,0.11$\pm$0.10  & \,\,0.48$\pm$0.17 \\
$C$            & \,\,0.0            & \,\,0.93$\pm$0.30 \\
$\sigma_{FIT}$ & \,\,-              & \,\,0.41$\pm$0.12 \\
$\alpha_{FIT}$ & \,\,-              & \,\,~~~~~~~~~1.57$\pm$0.081 (stat) \\
\hline
\end{tabular}
\linespread{1.0}
\caption{The best-fit parameters of equation (\ref{adjust_q3PACF}), for the BAO ($C\neq0$) and {\it non}-BAO ($C=0$) cases (see Figure \ref{fig1}), obtained through the $\chi^2$ statistics, 
equation (\ref{covmatrix}), using the covariance matrix shown in Figure \ref{fig2}.
}
\label{table1}
\end{table}

\subsection{Spectroscopic-$z$ Error}\label{sec4.3}

As shown by~\cite{Sanchez11}, the primary source of systematic errors in the 2PACF and, 
consequently, in the 3PACF, comes from the uncertainty in the measurement of the redshift, 
$z$, particularly large in the case of photometric redshift surveys 
with broad-band filters as
 the DES survey\footnote{\url{https://www.darkenergysurvey.org/}}. 
On the other hand, narrow-band filters from current~\citep{PAU} and forthcoming~\citep{JPAS} 
surveys deal with photo-$z$ errors that are competitive with spectro-$z$ errors. 

In the case of the sample studied here, the DR12Q catalogue from the SDSS, the data is 
spectroscopic and the estimation of $z$ is very precise, as described by~\cite{Paris}. 
The error associated to spectroscopic measurements is $\delta z=0.003$ 
~\citep[a $3\,\sigma$ error;][]{Laurent16} which we shall call spec-$z$ error.

\begin{figure}
\mbox{\hspace{-0.5cm}
\includegraphics[width=9.8cm, height=6.9cm]{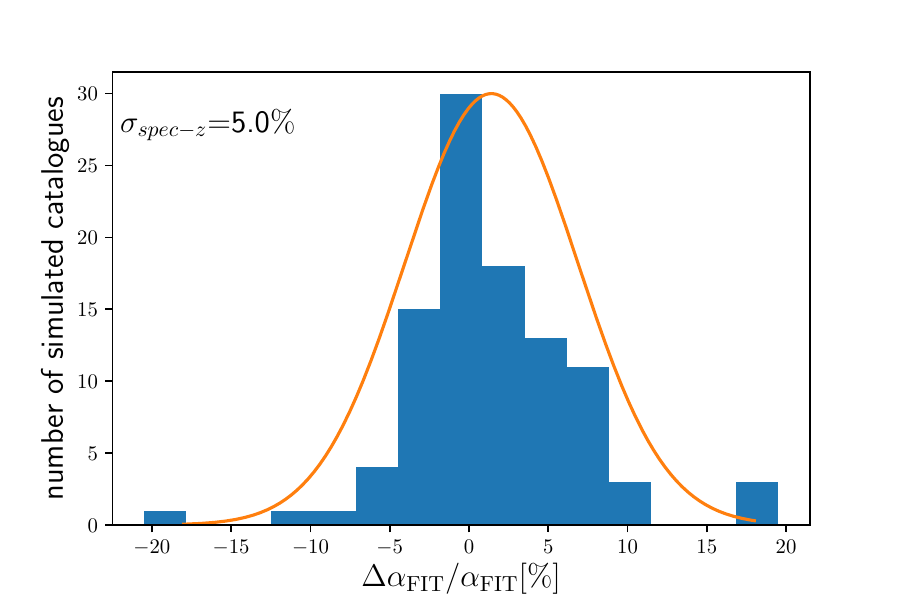}
}
\vspace{-0.4cm}
\caption{Histogram  of the difference $\Delta\alpha_{\mathrm{FIT}}$ between the $\alpha_{\mbox{\sc fit}}$ obtained from the 
quasars data compared to the values obtained from the simulated spec-$z$ samples. 
This spec-$z$ errors impact the measurements of the BAO signature as a systematic error 
with a relative amplitude of $5\,\%$.}
\label{fig4}
\end{figure}

To estimate the impact of a redshift uncertainty, $\delta z$, in our 
analyses, we shall perform a test. 
Consider that the redshift values, $z_i$, given in the DR12Q catalogue are the correct ones. 
For each $z_i$ we produce a simulated 
error  according to a Gaussian distribution with mean $z_i$ and standard deviation 
$(1 + z_i)\,\delta z$. 
We applied this methodology to generate 100 spec-$z$ quasar catalogues, where
a given quasar appears in each of these catalogues at a different
redshift, whose displacements from the correct values follow
such Gaussian distribution. 
In Figure~\ref{fig4} we show the difference between the 
$\alpha_{\mbox{\footnotesize\sc fit}}$ adjusted from the `true' quasar sample and the one 
obtained from each simulated spec-$z$ quasar sample. 
Then, the relative error associated with the spec-$z$ error is $5\,\%$ for the reduced 3PACF 
case, and $4\,\%$ for the 2PACF case. 
This means that the systematic error in the 3PACF BAO measurement is $0.08$ rad in $\alpha$. 
Other sources of systematic errors are the redshift space distortions and the projection effects. 
However, for the sample in study, their contribution is expected to be small \citep[see, 
e.g.,][]{Sanchez11}.

\subsection{Robustness of BAO signal and the Null test}

As observed by~\cite{Gaztanaga09}, a robustness test of the analyses in $n$-point correlation 
studies is made by confirming the BAO signature in the 2PACF and 3PACF individually. 
Once the signal is detected in both statistics one can consider that the BAO detection is robust. 
In our case, we have obtained the BAO signature using the 2PACF in EdC18, and in the 
present analyses we confirm it, with a good statistical significance, with the 3PACF.

\begin{figure}
\mbox{\hspace{-0.4cm}
\includegraphics[width=9.3cm, height=6.9cm]{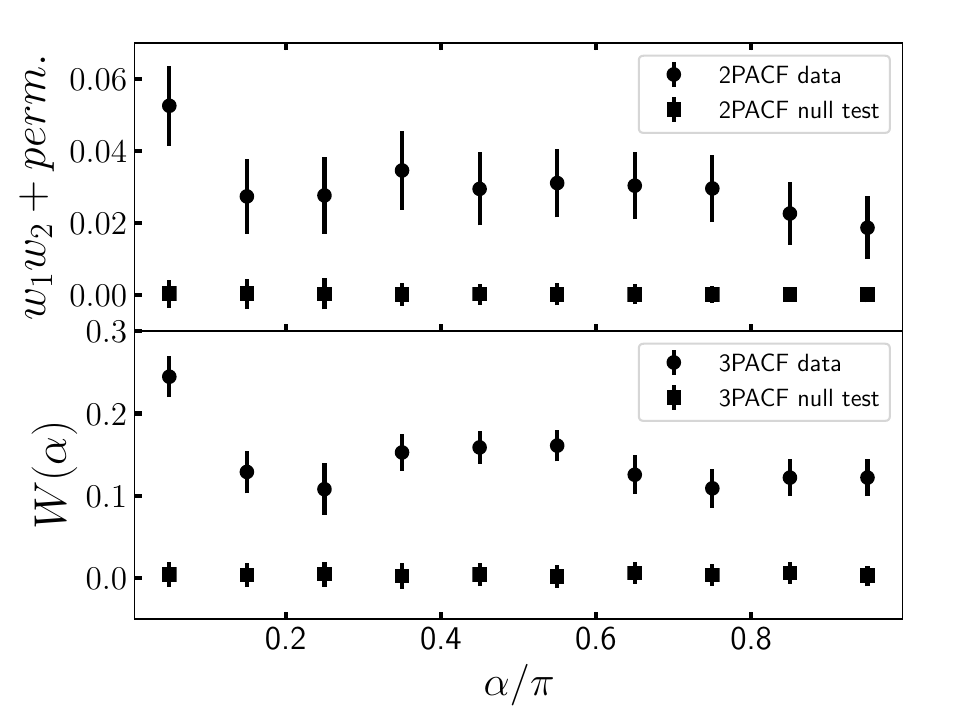}
}
\vspace{-0.4cm}
\caption{The combination of several 2PACF 
$w_{1}w_{2}+w_{1}w_{3}+w_{2}w_{3}$ (upper panel), and the 3PACF (bottom panel), 
where the data points (circles) correspond to the analyses of the quasars sample DR12. 
In both panels the data square symbols represent the null test, obtained by replacing the 
quasars data catalogue with a random catalogue, performing this operation 100 times, and then 
considering the mean and standard deviation for the data (squares) and error bars, respectively. 
} 
\label{fig5}
\end{figure}

We also perform the null test to investigate the behaviour of the 2PACF 
and 3PACF estimators when the data is replaced by a random catalogue; this procedure is 
repeated with several random catalogues to compute the average. 
For this, we generated 100 extra random samples (see section~\ref{sec2}), to replace the data 
100 times, and for each one we obtained the 2PACF and 3PACF, finally calculating the mean 
2PACF and the mean 3PACF. 
This procedure follows the same methodology as described in EdC18. 
The results are shown in the Figure~\ref{fig5}, where we present the $n$-point statistics 
($n=2,3$) calculated 
from the quasars data, that is: the 2PACF, $w(\theta)$, in the upper plot, and the 3PACF, 
$W(\alpha)$ in the bottom plot, both data represented by dot symbols. 
In the same panels of Figure~\ref{fig5} we show for comparison the results from to the null 
test analyses, represented by square symbols. All error bars are the standard deviation 
computed from the 100 datasets. 
As observed, for the null test the $w(\theta)$ and $W(\alpha)$ data points are zero, 
as expected, confirming that the random samples have no signature that could contaminate 
our results.

\subsection{Small Shifts Criterium}\label{ss-ssc}

This criterium is one more test to validate our results by examining if the signal observed in 
the 3PACF is not originated by statistical noise, an effect always present in the $n$-point 
correlation analyses. 

\begin{figure}
\mbox{\hspace{-0.5cm}
\includegraphics[width=9.8cm, height=6.9cm]{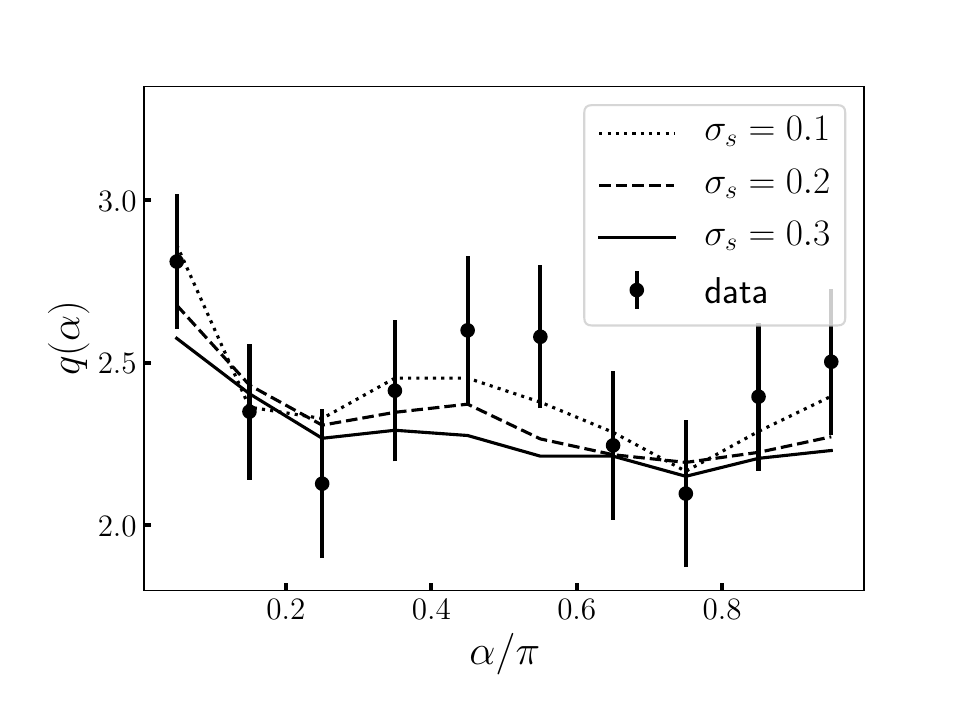}
}
\vspace{-0.4cm}
\caption{The reduced 3PACF, $q(\alpha)$, for the original quasars catalogue (dots), 
using the small shift criterium as described in the text (see the subsection~\ref{ss-ssc}). 
We have used $\sigma_s=0.1^\circ,0.2^\circ$, and $0.3^\circ$ (dotted, dashed and continuous lines, respectively) 
to perturb the original quasar positions. 
As observed, the BAO signature is very robust, clearly appearing still for shifts as intense as a 
Gaussian shift with $\sigma_s=0.2^\circ$ (the angular positions of the quasars are shifted at a 
maximum distance of $1.0^{\circ}$).} 
\label{fig6}
\end{figure}

To apply this test we have followed the same procedure used in 
EdC18~\citep[see also][]{Gabriela1}. 
We perturb the quasars original positions in the sky according to a Gaussian distribution, 
in three cases: considering its standard deviation as $\sigma_s = 0.1^\circ, 0.2^\circ$, 
and $0.3^\circ$. 
Geometrically, this process means that the quasars positions are randomly shifted in direction  
and with displacements of different sizes (following a Gaussian distribution). 
In Figure~\ref{fig6} we illustrate the effect of this procedure, where even for the intense shake 
corresponding to $\sigma_s = 0.2^\circ$ or maximum displacement of $1.0^{\circ}$, the BAO 
signal is highly suppressed but still survives exhibiting the robustness of our result. 
In the most severe case, $\sigma_s = 0.3^\circ$, a very small BAO signature could be there, 
but due to the error bars the result appears compatible with the absence of signal.

\subsection{Projection effect in the 3PACF}\label{proj-efec}

\noindent
To access the $\theta_{\mbox{\sc bao}}$ we need to correct the $\alpha_{\mbox{\sc fit}}$ 
with respect to the projection effect which produces a shift in the BAO bump 
position~\citep{Sanchez11}. 
For this, we convert $\zeta(S)$ into $W(\Theta)$  by using the 3PCF result provided by Pertubation Theory~\citep[see][]{fry1984, Frieman99, BarrigaGaztanaga2002, Gaztanaga09} and the relation 
\begin{eqnarray}\label{convertequation}
\begin{split}
	W(\theta_{12},\theta_{23},\theta_{13}) = \int dz_1\, \phi(z_1) \int dz_2\, \phi(z_2) \\
	\int dz_3\, \phi(z_3)\, \zeta(r_{12},r_{23},r_{13};\bar{z}) \, ,
\end{split}
\end{eqnarray}
where $\bar{z} = (z_1+z_2+z_3)/3$ and $\phi(z)$ is the redshift selection function 
normalized to unity within the shell of width $\delta{z}$. 
In the case of the 2PACF, we followed the procedure described in~\cite{Gabriela1,Edilson18}.  
As a reference model we used the $\Lambda$CDM model with cosmological parameters 
from Planck~\citep{PLA2018}. 

To calculate the projection effect for our data with $z \in [2.20,2.25]$, that is, in a shell of 
width $\delta{z}=0.05$, we evaluated the above relation in two cases: $\delta{z}=0.0$ and 
$\delta{z}=0.05$, and then calculated the relative difference, 
$\Delta \,\equiv\, ( \alpha_{\mbox{\sc fit}}|_{\delta{z}=0.05} 
- \alpha_{\mbox{\sc fit}}|_{\delta{z}=0.0} ) / \alpha_{\mbox{\sc fit}}|_{\delta{z}=0.0}$, 
in the BAO bump position. 
Thus, we obtain a relative difference of $\Delta = 1.12 \%$. 
Applying this shift to $\alpha_{\mbox{\sc bao}} = (1\,+\, \Delta) \alpha_{\mbox{\sc fit}}$, 
we obtain $\alpha_{\mbox{\sc bao}} = 1.59\,$rad.

\subsection{Validation of the results via fiducial cosmology}

Finally, we shall test the validity of our results assuming a fiducial cosmology. 
In fact, it is important to evaluate if the use of an empirical parametrization, 
as given by equation~(\ref{adjust_q3PACF}), could bias our result.
To do this, we use the theoretical realisation of the reduced 3PACF \citep[applying the same procedure described by][]{BarrigaGaztanaga2002}, considering as fiducial cosmology the flat $\Lambda$CDM, with $(\Omega_m, h, \Omega_b, \sigma_8. n_s) = (0.31, 0.7, 0.059, 0.8, 0.97)$, and using a non linear power spectrum derived from Perturbation Theory. 
We follow a local biasing model with  \citep{Fry93,Frieman94,desjacques18}
\begin{eqnarray}
\delta_Q \,=\, \sum_{k=0}^{\infty} \, \frac{b_k}{k!} \, \delta_m^k \, ,
\end{eqnarray}
where $\delta_i$ is the density contrast, for quasar, $i=Q$, and for matter, $i=m$. 
This way, in the leading order we have, for the reduced 3PACF, $q_Q = (1/b_1)q_m + b_2/b_1^2$, 
where $b_1$ is the usual linear (local) bias, used here as the effective bias parameter $b_{eff} = 4.25$ for quasars \citep{Laurent16}. 
Since the non-local bias term contributes only by shifting the $q_Q$ curve, with no effect in its 
shape, we consider $b_2 = 0$ \citep{Frieman99}. 

\begin{figure}
\mbox{\hspace{-0.5cm}
\includegraphics[width=9.8cm, height=6.9cm]{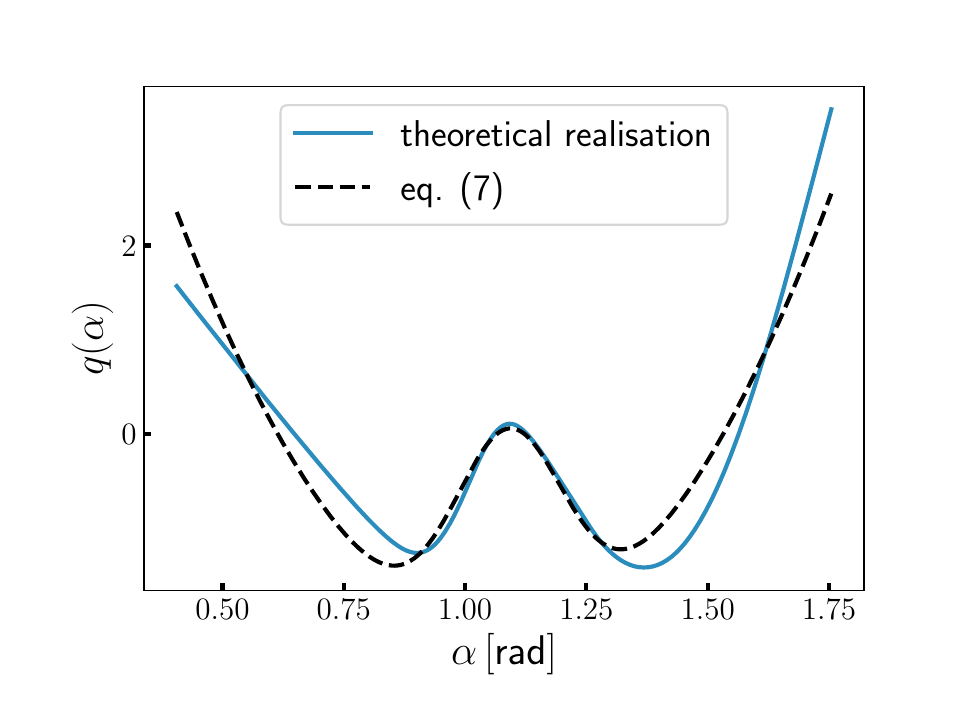}
}
\vspace{-0.4cm}
\caption{Theoretical calculation of the reduced 3PACF (continuous line) produced according to \citet{BarrigaGaztanaga2002} using the CDM power spectrum solution \citep[derived from Perturbation Theory;][]{fry1984}, including the effective bias parameter for quasars (see text for details). 
The dashed line represents an illustrative example of the fitting procedure using equation (\ref{adjust_q3PACF}). 
For comparative purposes, we plotted the curves around the BAO bump.
The precision and accuracy to recover the BAO signal position was also examined and the 
results are displayed in the Figure \ref{fig8}. 
}
\label{fig7}
\end{figure}

In addition, following \citet{Sanchez11, Crocce11}, we model the selection function as 
\begin{eqnarray}
\phi(z) \,=\, \frac{dN_Q}{dz} \, W(z) \, ,
\end{eqnarray}
where $W(z)$ is the window function encoding our redshift cuts, $z \in [2.20, 2.25]$ (i.e., $W=1$ inside the shell and $W=0$ outside the shell). 
The term $dN_Q/dz$ corresponds to the distribution of quasars as a function of the redshift, 
chosen to be the Gaussian curve best-fitted to the DR12Q distribution, in the range $1.1 < z < 4.0$. 
The reduced 3PACF resulting from such theoretical calculation is shown as a continuous line in Figure \ref{fig7}. 
Notice that this theoretical curve is obtained by fixing the angular distances at $\theta_1 = 1.0^\circ$ and $\theta_2 = 1.5^\circ$, and the range $\alpha = [0,\pi]$, i.e., the same values used in the data analyses.

\begin{figure}
\mbox{\hspace{-0.5cm}
\includegraphics[width=9.8cm, height=6.9cm]{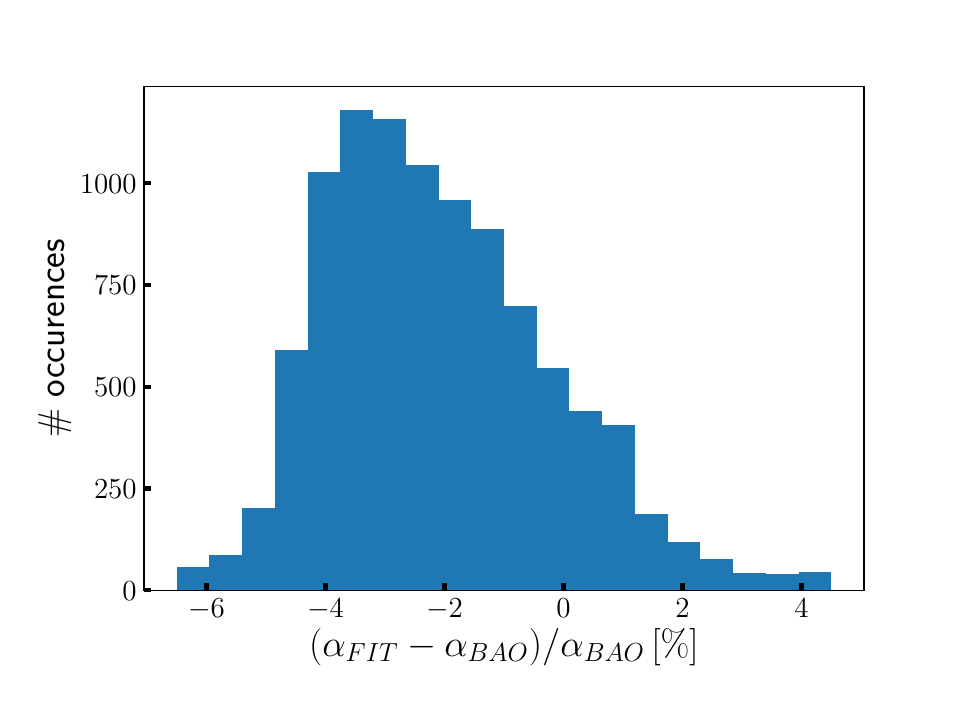}
}
\vspace{-0.4cm}
\caption{Performance test for recovering the $\alpha_{BAO}$. 
Histogram of the relative difference between the $\alpha_{BAO}$ (input value) and 
the best-fitting $\alpha_{FIT}$ value obtained by using the equation~(\ref{adjust_q3PACF}) 
for $10,000$ synthetic realisations of $q(\alpha)$. This analysis reveals that the mean 
value of the differences has a 2.1\% deviation with respect to the input value, which 
represents an error of $\sigma_{param} = 0.033$ rad.}
\label{fig8}
\end{figure}

Then, to validate the performance of equation (\ref{adjust_q3PACF}) in correctly fitting the data, 
we repeated the same procedure of generating 10,000 synthetic realizations
described in section \ref{bin-size}, but using the theoretical $q(\alpha)$ (continuous line in 
Figure \ref{fig7}) as the true one.
Figure \ref{fig8} shows the histogram of the relative difference between the input (true) 
value $\alpha_{BAO}$ and the $\alpha_{FIT}$ estimates obtained by fitting the 
equation~(\ref{adjust_q3PACF}) to each of these realizations (see Figure \ref{fig7} for an 
illustrative example of this fitting procedure). 
The mean value of these differences has a $2.1\%$ deviation with respect to the input value, 
which represents an error of $\sigma_{param} = 0.033$ rad. 
In fact, this error accounts for only a small fraction of the systematic error, 
whose main contribution comes from the spectroscopic error. 
Adding both contributions in quadrature we have $\sigma_{sys} = 0.087$ rad.
Finally, we use equation~(\ref{eq6}) to find $\theta_{\mbox{\sc bao}} = 1.82^\circ\pm 0.21^{\circ}$, 
considering a combined computation using both the statistical and the systematic errors, 
as done by~\cite{Carnero11}, and the errors in $\theta_1$ and $\theta_2$.

\section{Conclusions and Final remarks}\label{sec5}

The clustering of matter structures in the Universe is currently probed with large deep surveys 
by the $n$-point correlation function. 
In a previous work, EdC18, we used the angular version of the 2-point statistic, the 2PACF, 
to study the BAO phenomenon in the DR12 quasars catalogue from the SDSS, 
with $z \in [2.20, 2.25]$ considering quasars located in the north Galactic hemisphere, detecting 
the transversal BAO signal at 
$\theta_{\mbox{\sc bao}}^{\mbox{\footnotesize\sc 2PACF}} = 1.77^{\circ} \pm 0.31^{\circ}$. 
Here we also studied the BAO features in the SDSS DR12 quasars catalogue, in the 
same redshift bin as the above analyses but using the angular version of the 3-point statistic, the 
3PACF, considering data from the north and south Galactic hemispheres. 
We detect a transversal BAO signal with statistical significance, of $2.9\,\sigma$, at 
$\theta_{\mbox{\sc bao}}^{\mbox{\footnotesize\sc 3PACF}} 
= 1.82^{\circ} \pm 0.21^{\circ}$, 
in excellent agreement with the measurement done with the 2PACF~\citep{Edilson18}, 
successfully confirming this BAO signature for quasars at the mean redshift $\bar{z}=2.225$.

Additionally, we also performed diverse robustness tests to confirm several steps of our 
procedure to find the BAO signature with the reduced 3PACF in these quasars data. 
To estimate the error bars and the significance of our results we have used a sample of 200 
quasar mocks. 
For each mock, we extracted the information about the 2PACF and the 3PACF and finally calculated 
the reduced 3PACF, $q(\alpha)$. 
The covariance matrix for each case was estimated using the procedure explained in the 
subsection~\ref{ss-covmatrix}.  

The significance of the result was accessed comparing the parametrization given in the 
equation~(\ref{adjust_q3PACF}) with and without BAO signal and using the inverse of the 
covariance matrix coming from the mocks. 
Finally, the successful result from the null-test guarantees that the random samples have no 
signature that could contaminate our results. 

\section*{Acknowledgements}
The authors acknowledge insightful discussions with Thiago Pereira and Eus\'ebio S\'anchez. 
EdC acknowledges the PROPG-CAPES/FAPEAM program. 
AB, HX, and CN acknowledge the support of the Brazilian agencies 
CNPq, FAPESP, and FAPERJ, respectively. 
We also thank the CAPES PVE project 88881.064966/2014-01, within the {\em Science 
without Borders Program}.


\end{document}